\documentclass[aps,pra,twocolumn,amsmath,amssymb,nofootinbib,showpacs,superscriptaddress]{revtex4-1}

\usepackage[english]{babel}
\usepackage{latexsym}
\usepackage{graphics}
\usepackage{graphicx}
\usepackage{epsfig}
\usepackage{color}
\usepackage{bm}
\usepackage{amsmath}
\usepackage{amssymb}
\usepackage{amsthm}
\usepackage{dcolumn}
\usepackage{physics}
\usepackage{bm}
\usepackage{float}
\usepackage{hyperref}
\usepackage{epstopdf}
\usepackage{cleveref}
\usepackage[svgnames]{xcolor}
\usepackage{enumerate}
\usepackage[normalem]{ulem}
\usepackage{braket}

\hypersetup{hidelinks,
            colorlinks=true,
            allcolors=DarkBlue}

\usepackage{lineno}

\begin{document}
	
	\preprint{APS/123-QED}
	
	\title{Learning entanglement breakdown as a phase transition by confusion}

	\author{M.A. Gavreev}
	\affiliation{Russian Quantum Center, Skolkovo, Moscow 143025, Russia}
	\affiliation{Moscow Institute of Physics and Technology, Dolgoprudny, Moscow Region 141700, Russia} 
	
	\author{A.S. Mastiukova}
	\affiliation{Russian Quantum Center, Skolkovo, Moscow 143025, Russia}
	\affiliation{Moscow Institute of Physics and Technology, Dolgoprudny, Moscow Region 141700, Russia} 
	\affiliation{Schaffhausen Institute of Technology, Schaffhausen 8200, Switzerland} 
	
	\author{E.O. Kiktenko}
	\affiliation{Russian Quantum Center, Skolkovo, Moscow 143025, Russia}
	\affiliation{Moscow Institute of Physics and Technology, Dolgoprudny, Moscow Region 141700, Russia} 
	\affiliation{Department of Mathematical Methods for Quantum Technologies, Steklov Mathematical Institute of Russian Academy of Sciences, Moscow 119991, Russia}
	
	\author{A.K. Fedorov}
	\affiliation{Russian Quantum Center, Skolkovo, Moscow 143025, Russia}
	\affiliation{Schaffhausen Institute of Technology, Schaffhausen 8200, Switzerland} 
	
	\date{\today}
	\begin{abstract}
	Quantum technologies require methods for preparing and manipulating entangled multiparticle states. 
	However, the problem of determining whether a given quantum state is entangled or separable is known to be an NP-hard problem in general, and even the task of detecting entanglement breakdown for a given class of quantum states is difficult. 
	In this work, we develop an approach for revealing entanglement breakdown using a machine learning technique, which is known as ‘learning by confusion’.
	We consider a family of quantum states, which is parameterized such that there is a single critical value dividing states within this family into separate and entangled. 
	We demonstrate the ‘learning by confusion’ scheme allows us to determine the critical value. 
	Specifically, we study the performance of the method for the two-qubit, two-qutrit, and two-ququart entangled state.
	In addition, we investigate the properties of the local depolarization and the generalized amplitude damping channel in the framework of the confusion scheme. 
	Within our approach and setting the parameterization of special trajectories, we obtain an entanglement-breakdown ‘phase diagram’ of a quantum channel, 
	which indicates regions of entangled (separable) states and the entanglement-breakdown region. 
	Then we extend the way of using the ‘learning by confusion’ scheme for recognizing whether an arbitrary given state is entangled or separable. 
	We show that the developed method provides correct answers for a variety of states, including entangled states with positive partial transpose (PPT). 
	We also present a more practical version of the method, which is suitable for studying entanglement breakdown in noisy intermediate-scale quantum (NISQ) devices. 
	We demonstrate its performance using an available cloud-based IBM quantum processor.
	\end{abstract}
	
\maketitle
	
\section{Introduction}
	
Machine learning is considered as a useful tool for analyzing patterns in data, which makes it attractive for quantum physics research~\cite{Dunjko2018}. 
Recent results are in particular related to exploring machine learning methods for studying states of matter and phase transitions~\cite{Wang2016,Wetzel2017,Melko2017,Melko2018,Kim2019,Broecker2016,Schindler2017,Chng2017,Nieuwenburg2017,Ringel2018,Beach2018,Greitemann2018,Knap2019,Rem2019,Liu2019,Dong2019,Kharkov2020}, 
efficient representations of many-body quantum states~\cite{Troyer2017,Glasser2018,Lu2018,Troyer2018,Tiunov2020}, and controlling experiments~\cite{Sriarunothai2018,Zhang2018,Monroe2013,Rey2017}. 
A specific task, which is of interest both from the side of fundamental aspects of quantum theory and  potential applications, is related to the classification between entangled or separable quantum states. 
It is known that the general problem of determining whether a given quantum state is entangled or separable is computationally intensive, and it is known to be NP-hard~\cite{Gurvits2003}. 
There are exact known results for entanglement measures, but they are mostly limited to low-dimensional quantum systems~\cite{Peres1996,Horodetski1996,Lutkenhaus2006}. 
The well-known criterion is the positive partial transposition (PPT) criterion, which says that a separable state must have a PPT~\cite{Peres1996,Horodetski1996}. 
However, this condition is necessary and sufficient only for the case of the dimensions of the subsystems $A$ and $B$ that satisfy the inequality $d_Ad_B \leq 6$. Thus, this criterion does not work, for example, for two-qutrit quantum systems. Another criterion is the $N$-symmetric extension hierarchy, which is currently one of the most powerful  approach~\cite{Plenio2009}. The downside of this criterion, however, is its exponentially growing with $N$ computational cost.
Necessary and sufficient methods have also been developed \cite{Horodetski1996, Wu2000, Terhal2000}, but they are difficult for practical use or have restrictions \cite{Li2018}.

The problem of detecting entanglement has received significant attention in the context of machine learning~\cite{Zeng2018,Ma2018,Gao2018,Wang2019,Chen2019,
Giraud2019,Guo2019,Jaffali2019,Deng2019,Bharti2020}. 
The proposed methods are quite efficient and provide high enough classification accuracy. However, such methods are mostly based on known properties of entanglement or require a complex set of labeled experimental data. In this regard, unsupervised (semi-supervised) machine learning methods are of particular interest, since they allow building an agnostic approach to study the entanglement phenomenon.

\begin{figure}[]
\includegraphics[width=0.9\linewidth]{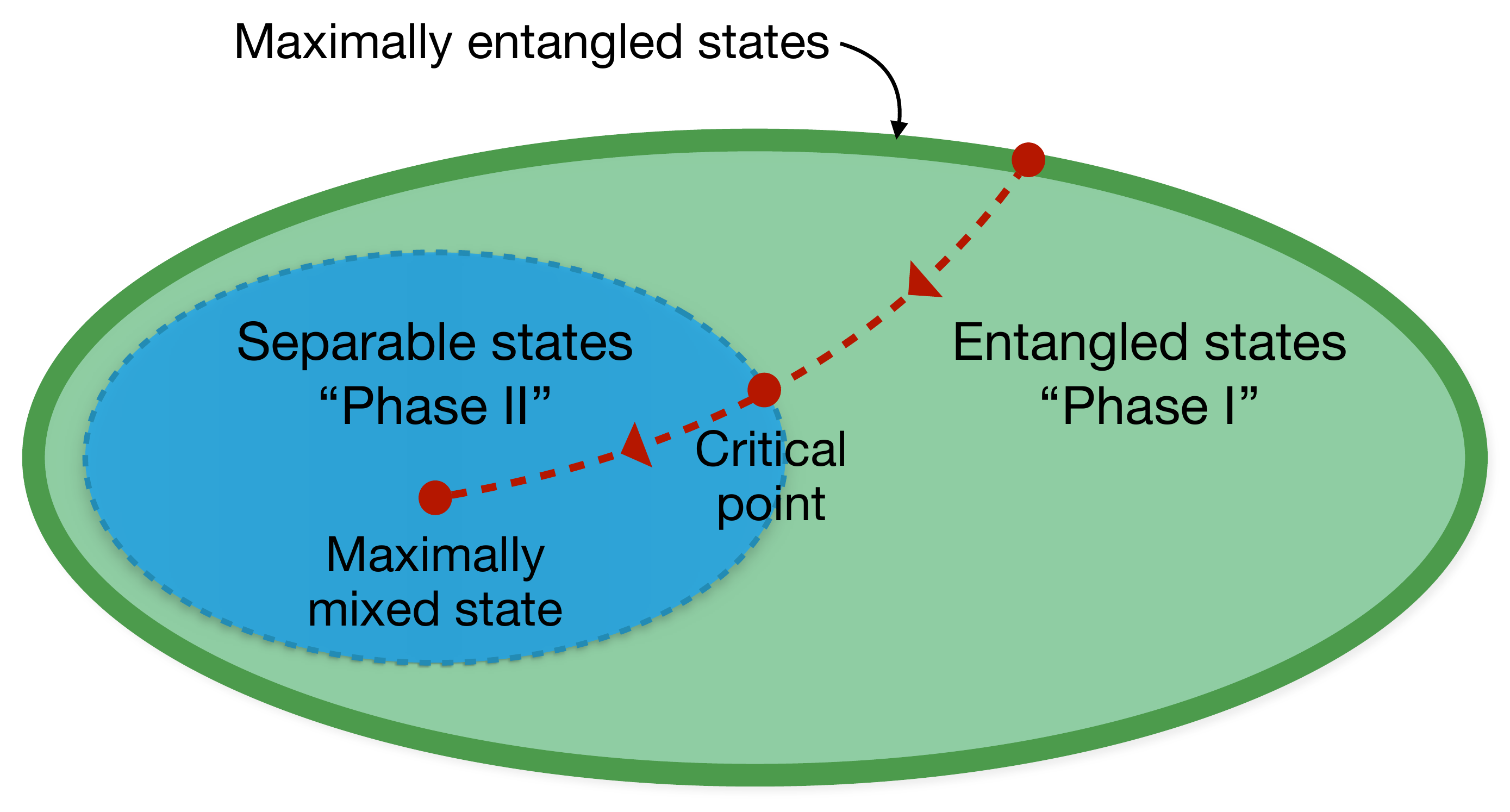}
\caption{ The confusion scheme for learning entanglement: 
First, we construct a path in the state space of a bipartite system that connects a maximally entangled state with a certainly separable state, e.g. the maximally mixed state.
Then we apply the confusion scheme to search for a phase transition point on this path and expect that found critical point is located at the border of entangled and separable states.}
\label{fig:EntSepDepolarization}
\end{figure}

Among a number of successful semi-supervised machine learning methods for analyzing quantum systems, the `learning by confusion' scheme~\cite{Nieuwenburg2017} deserves special attention due to its simplicity. 
This method uses the dynamical reassignment of the class labels concerning a given value of the parameter that is responsible for the transition. 
As an output of the scheme, a characteristic W shape of the performance of the recognition function is expected, with the middle peak corresponding to the transition parameter.
The `learning by confusion' scheme has been originally approached for the topological phase transition in the Kitaev chain, the thermal phase transition in the classical Ising model, 
and the many-body-localization transition in a disordered quantum spin chain~\cite{Nieuwenburg2017}, 
and then for applied for studying vortices at the Kosterlitz-Thouless transition~\cite{Melko2018}, double phase transitions and quasi-long-range order~\cite{Kim2019},
transitions between regular and chaotic states of spin chains~\cite{Kharkov2020}, and phase transitions in nonlinear polariton lattices~\cite{Kyriienko2022}, and others \cite{Bohrdt2021, Greplova2020}.

\begin{figure*}[t]
\includegraphics[width=1\linewidth]{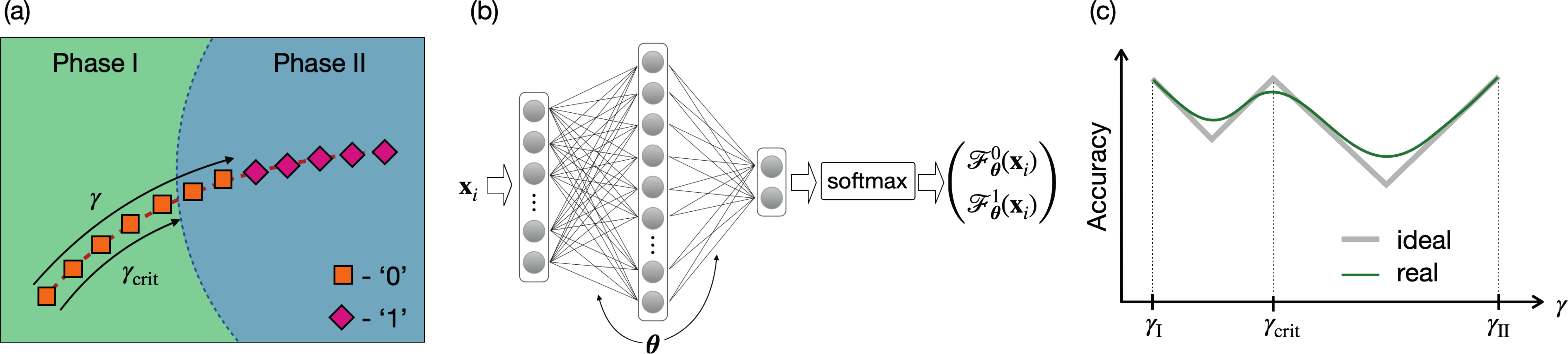}
\caption{The general scheme of the `learning by confusion' method is illustrated.
In (a) the idea of preparing and labeling an input data set is shown. 
Data points correspond to physical system states taken from a path connecting regions of phase I and phase II.
Each point is labeled with `0' or `1' according to the confusion parameter $\gamma$.
In (b) the application of a FFNN is shown.
A datapoint ${\bf x}_i$ is fed as an signal to input layer of the FFNN.
The signal from two output neurons is then transformed to a binomial probability distribution $({\cal F}_{\boldsymbol\theta}^0({\bf x}_i),{\cal F}_{\boldsymbol\theta}^1({\bf x}_i))$ using a standard softmax approach (we use a neural network with one hidden layer).
In (c) the ideal and real behavior of the FFNN performance as a function of $\gamma$ is shown.
The middle peak of the observed W shape curve corresponds to the revealed critical point.
}
\label{fig:confusion}
\end{figure*}

In this work, we consider several strategies of applying the `learning by confusion' scheme in the context of the investigation of entanglement breakdown i.e. transition from the subspace of entangled states to the subspace of separable states while moving along a given path.
We consider a family of quantum states, which is parametrized such that there is a single critical value dividing states within this family on separate and entangled. 
We demonstrate the `learning by confusion' scheme allows for determining the corresponding critical value of the parameter.
We first present an approach that uses an action of a depolarizing channel on a maximally entangled state (see Fig.~\ref{fig:EntSepDepolarization}).
Within this method, we reveal entanglement breakdown and obtain corresponding W shapes for two-qubit two-qutrit, and two-ququart cases.
We then use the `learning by confusion' scheme for the characterization of entanglement breakdown quantum channels and calculate their properties as `phase diagrams',
whose borders separates entangled and separable subsets of states.
Second, we extend this method to the analysis of arbitrary states using a state-specific transformation to construct a path connecting the maximally entangled and maximally mixed states.
Finally, we present a practical version of our method, which is suitable for application to studies entanglement breakdown in quantum circuits using a cloud-base IBM quantum processor.

We note that a specific aspect of the application of machine learning methods to quantum data is the fact that they are not typically described in terms of regular (nonnegative and normalized) probability distributions~\cite{Melko2019}. 
A possible way to use probability distribution for the description of quantum states is to employ informationally complete (IC) positive-operator valued measures (POVMs), 
and their symmetric versions (SIC-POVMs)~\cite{Stacey2017,Stacey2017-2,Kiktenko2020,Kiktenko20202}. 
Here we use SIC-POVMs giving probability representations, which completely describe quantum states so that corresponding probability distribution can be used as input data for machine learning schemes. 

Our work is organized as follows. 
In Sec.~\ref{sec:Confusion}, we describe the `learning by confusion' scheme and its adoption for learning entanglement. 
In Sec.~\ref{sec:W}, we present a method based on the action on maximally entangled state and provide results for two-qubit, two-qutrit and two-ququart quantum states with providing W shapes.
We also present `phase diagrams' for entanglement-breakdown quantum channels that are associated with our method. 
In Sec.~\ref{sec:EntClassification}, we describe the data set preparation technique that allows us to study the entanglement of an arbitrary quantum state.
In Sec.~\ref{sec:IBM}, we show the application of our approach using for data collected from the superconducting $5$-qubits quantum processor IBM Athens. 
We summarize our results and conclude in Sec.~\ref{sec:Conclusion}.

\section{Confusion scheme for learning entanglement}\label{sec:Confusion}

Here we first introduce the `learning by confusion' scheme, which has been originally proposed as an approach for finding phase transitions~\cite{Nieuwenburg2017}. 
Then we describe data set preparation for the usage of the scheme for studying entanglement. 

\subsection{`Learning by confusion' scheme}

The `learning by confusion' scheme is based on the analyzing performance of a feed-forward neural network (FFNN) after it is trained with deliberately incorrectly labeled data. 
The scheme works as follows.
Suppose we are dealing with a physical system, whose state space (or region of the state space of interest) ${\cal X}$ consists of two areas: the one corresponded to `phase I' and another to `phase II' (see Fig.~\ref{fig:confusion}).
Let us consider a curve in ${\cal X}$, parametrized by a real parameter ${\gamma} \in [\gamma_{\rm I}, \gamma_{\rm II}]$ in such a way that is starts in the phase I ($\gamma=\gamma_{\rm I}$) and ends in phase II ($\gamma=\gamma_{\rm II}$).
Assuming that the curve crosses a phase transition border once, the problem is to find an unknown critical point $\gamma_{\rm crit}\in(\gamma_{\rm I}, \gamma_{\rm II})$ where the phase transition occurs.

We are going to solve this problem by processing a dataset 
\begin{equation}
	{\sf data}=\{ ({\bf x}_i, \gamma_i) \}_{i=1}^{N}
\end{equation}
of $N$ data points ${\bf x}_i\in {\cal X}$ distributed `uniformly' on the curve: each point ${\bf x}_i$ corresponds to $\gamma=\gamma_i$.
To apply the FFNN to the data, ${\sf data}$ is randomly splitted into two uneven parts: the training datatset ${\sf data}_{\rm train}$ used for tuning parameters of the FFNN , 
and the test dataset  ${\sf data}_{\rm test}$ employed for the performance evaluation.

The core feature of the confusion scheme is the labeling of datasets.
Let us pick up some arbitrary point $\gamma\in[\gamma_{\rm I}, \gamma_{\rm II}]$, and pretend that it is the critical point of the phase transition.
In what follows, we refer to $\gamma$ as a confusion parameter.
According to the choice of $\gamma$, we label all data points ${\bf x}_i$ in ${\sf data}_{\rm train}$  and ${\sf data}_{\rm test}$ having $\gamma_i\leq \gamma$ with label `0', and all remaining points with label `1' [see Fig.~\ref{fig:confusion}(a)].
More formally, we prepare datasets
\begin{equation} \label{eq:datasets-prep}
	\begin{aligned}
		X_{\rm train/test}^{(0)}(\gamma) &= \{{\bf x}_i: (x_i,\gamma_i)\in {\sf data}_{\rm train/test}, \gamma_i\leq\gamma\},\\
		X_{\rm train/test}^{(1)}(\gamma) &= \{{\bf x}_i: (x_i,\gamma_i)\in {\sf data}_{\rm train/test}, \gamma_i>\gamma\}.
	\end{aligned}
\end{equation}
The obtained datasets $\{X_{\rm train}^{(r)}(\gamma)\}_{r=0,1}$ are then used to `confuse' the neural network in the training process of minimizing loss function
\begin{equation}
	{\cal J}(\boldsymbol\theta) = -\sum_{r=0,1}\sum_{{\bf x}_i\in X_{\rm train}^{(r)}(\gamma)} \log {\cal F} ^{1-r}_{\boldsymbol\theta}({\bf x}_i),
\end{equation}
where ${\cal F}^j_{\boldsymbol\theta}({\bf x}_i)\in(0,1)$ is an output probability obtained from FFNN, specified by weight coefficients $\boldsymbol\theta$, that the input data point ${\bf x}_i$ has label $j\in\{0,1\}$ [see Fig.~\ref{fig:confusion}(b)].
The resulting solution --- tuned weight coefficients $\boldsymbol\theta_{\gamma}$ --- are then used for evaluating performance (accuracy) of the neural network ${\cal P}(\gamma)$ given by a fraction of `correct answers' on test sets:
\begin{multline}
	{\cal P}(\gamma) = \frac{1}{|{\sf data}_{\rm test}|} \left( \sum_{{\bf x}_i\in X_{\rm test}^{(0)}(\gamma)} (1-\widehat{\cal F}_{\boldsymbol\theta_{\gamma}}({\bf x}_i)) \right.\\+
	\left.\sum_{{\bf x}_i\in X_{\rm test}^{(1)}(\gamma)} \widehat{\cal F}_{\boldsymbol\theta_{\gamma}}({\bf x}_i) \right)\in(0,1),
\end{multline}
where $|\cdot|$ stands for the number of elements in the set and
\begin{equation}
	\widehat{\cal F}_{\boldsymbol\theta_{\gamma}}({\bf x}_i)=
	\begin{cases}
	0 &\text{if } {\cal F}^0_{\boldsymbol\theta_{\gamma}}({\bf x}_i) \geq 1/2\\
	1 &\text{otherwise.}
	\end{cases}
\end{equation}
and the `hard decision' output of the FFNN.

Let us consider how the confusion parameter $\gamma$ affects the performance of the FFNN.
First of all, we expect that ${\cal P}(\gamma_{\rm I})={\cal P}(\gamma_{\rm II})=1$ since in these cases the training and test sets consist only of points with definite label.
Then, one can also expect a high performance ${\cal P}(\gamma_{\rm crit})\approx 1$ near the true critical point of the phase transitions $\gamma\approx \gamma_{\rm crit}$.
Meanwhile, in the intermediate regions $(\gamma_{\rm I}, \gamma_{\rm crit})$ and $(\gamma_{\rm crit}, \gamma_{\rm II})$ it is expected that the performance function ${\cal P}(\gamma)$ should drop down due to the fact of confusion. 
In the result, we expect to have ${\cal P}(\gamma)$ in a form of W shape, in the ideal case, where all data points belonging to particular phase are indistinguishable for the neural network, given by
\begin{equation}\label{eq:performance}
        \mathcal{P}(\gamma) =
        \begin{cases}
        1 - \frac{{\rm min}(\gamma_{\rm crit} - \gamma , \gamma -\gamma_{\rm I})}{\gamma_{\rm II}-\gamma_{\rm I}} &
        \text{for $\gamma_{\rm I} < \gamma < \gamma_{\rm crit}$},\\
         1 - \frac{{\rm min}( \gamma -\gamma_{\rm crit},\gamma_{\rm II} - \gamma)}{\gamma_{\rm II}-\gamma_{\rm I}}, &
         \text{for $\gamma_{\rm crit} <  \gamma < \gamma_{\rm II}$},
        \end{cases}
\end{equation}
shown in Fig.~\ref{fig:confusion}(c).
In the real experimental settings, where the neural network may manage to identify some structure within each phase, W shapes commonly have a smoother form. 
However, for a properly designed neural networks there appear an apparent middle peak in $\mathcal{P}(\gamma)$ 
that accurately coincides with the true critical point $\gamma_{\rm crit}$~\cite{Nieuwenburg2017,Melko2018,Kim2019,Kharkov2020,Kyriienko2022}.

\subsection{Data preparation for learning entanglement} \label{sec:dataprep}

Since this scheme is heuristic-based there are no limitations on the type of data that can be used for the confusion scheme. 
In particular, we can use entangled and separable states of a parametric family of states and by using the scheme it is possible to find the point corresponding to the transition between entangled and separable states. 
Below we describe in detail how this goal can be achieved. 

The main objects of study are bipartite quantum states $\rho_{AB}\equiv\rho$ of two finite dimensional particles $A$ and $B$.
Let $d_{A}$ and $d_B$ be dimensions of particle $A$ and $B$, correspondingly.
Then $\rho$ is given by $d_Ad_B\times d_Ad_B$ Hermitian positive semi-definite unit-trace matrix.
Remember, that $\rho$ is called separable if it can be written in the form
\begin{equation} \label{eq:separable}
	\rho=\sum_i p_i \rho_A^{(i)} \otimes \rho_B^{(i)}, 
\end{equation}
where  $p_i,\rho_A^{(i)},\rho_B^{(i)} \geq 0$, ${\rm Tr}\rho_A^{(i)}={\rm Tr}\rho_B^{(i)}=1$, and $\sum_i p_i=1$.
Otherwise, $\rho$ is entangled.
We note that although in the general case determining whether $\rho$ is entangled or separable is a difficult computational problem, 
we can easily construct entangled states in the form $\rho=\ket{\phi}\bra{\phi}$ for some $\ket{\phi}$ with Schmidt rank larger than 1, and separable state in form~\eqref{eq:separable}.

For training neural networks, we represent density matrix $\rho$ in the form of $d_A^2d_B^2$-dimensional probability vector (probability distribution):
\begin{equation}
	\rho \leftrightarrow \vec{p}=\begin{bmatrix} p^{1,1} \\ \vdots \\ p^{d_A^2,d_B^2}  \end{bmatrix},
\end{equation}
where probabilities $p^{i,j}$ are obtained using symmetric informationally complete positive operator-valued measures (SIC-POVM) in the state space of particles $A$ and $B$
\begin{equation}
	\vec{p}^{~i,j}={\rm Tr}[\rho E_{A}^i\otimes E_{B}^j].
\end{equation}
Here $\{E_A^i\}_{i=1}^{d_A^2}$ and $\{E_B^j\}_{j=1}^{d_B^2}$ are sets of SIC-POVM rank-one effects, satisfying conditions
\begin{eqnarray}
	&E^{j}_{A(B)} \geq 0, \quad \sum_j E^{j}_{A(B)} = {\bf 1}_{A(B)},\\
	&\mathrm{Tr} \left[ E^{j}_{A(B)}E^{k}_{A(B)} \right]=\frac{d_{A(B)}\delta_{j,k}+1}{(d_{A(B)}+1)d_{A(B)}^2},
\end{eqnarray}
where ${\bf 1}_{A(B)}$ is the identity operator in the state space of particle $A(B)$ and $\delta_{j,k}$ stands for the Kronecker symbol.
Though an existence of SIC-POVMs is shown only for a limited list of dimensions~(for a review, see Refs.~\cite{Stacey2017,Stacey2017-2}), 
they can be constructed for systems we consider in the present work (for details, see Appendix~\ref{apx:SICPOVM}).

The resulting dataset has the following form:
\begin{equation} \label{eq:Dataset}
	{\sf data}=\{ \vec{p}_i, \gamma_i) \}_{i=1}^{N},
\end{equation}
where probability vectors $\vec{p}_i$ corresponds to a bipartite density matrix $\rho_i$ located (according to the value of $\gamma_i$) on the curve connecting region of entangled and and separable states.

\section{W shapes for entanglement breakdown}\label{sec:W}

Here we study entanglement-breakdown by constructing W shapes appearing in the process of passing maximally entangled states through quantum channels.
First, we justify the confusion scheme by considering a depolarizing channel acting on two-qubit, two-qutrit, and two-ququart states as a whole.
Then we study `phase diagrams' for entanglement-breakdown within the action of two quantum channels, acting separately on particles $A$ and $B$ of the maximally entangled two-qubit state.

\subsection{Depolarizing of maximally entangled states}\label{subsec:Dep}

The main idea of our method is to construct a `trajectory' in the space of quantum states, which allows passing through points between the space of maximally entangled states and the space of separable states (see Fig.~\ref{fig:EntSepDepolarization}).
This kind of trajectory can be obtained by applying a decoherence channel to a certain maximally entangled state.
As an example, we consider a depolarizing channel acting on two particles of the same dimension $d_A=d_B=d$:
\begin{equation}\label{eq:Depolarization}
	\Phi_{\alpha}^d[\rho]:=  \alpha \rho + (1 - \alpha)
	\frac{\textbf{1}\otimes \textbf{1}}{d^2} \cdot \textrm{Tr} [\rho],
\end{equation}   
where depolarization parameter $\alpha \in \left[ -(d^2-1)^{-1}, 1 \right]$ and $\textbf{1}=\textbf{1}_A=\textbf{1}_B$.
In what follows, $\alpha$ serves as the confusion confusion parameter.

\begin{figure}[]
\includegraphics[width=0.8\linewidth]{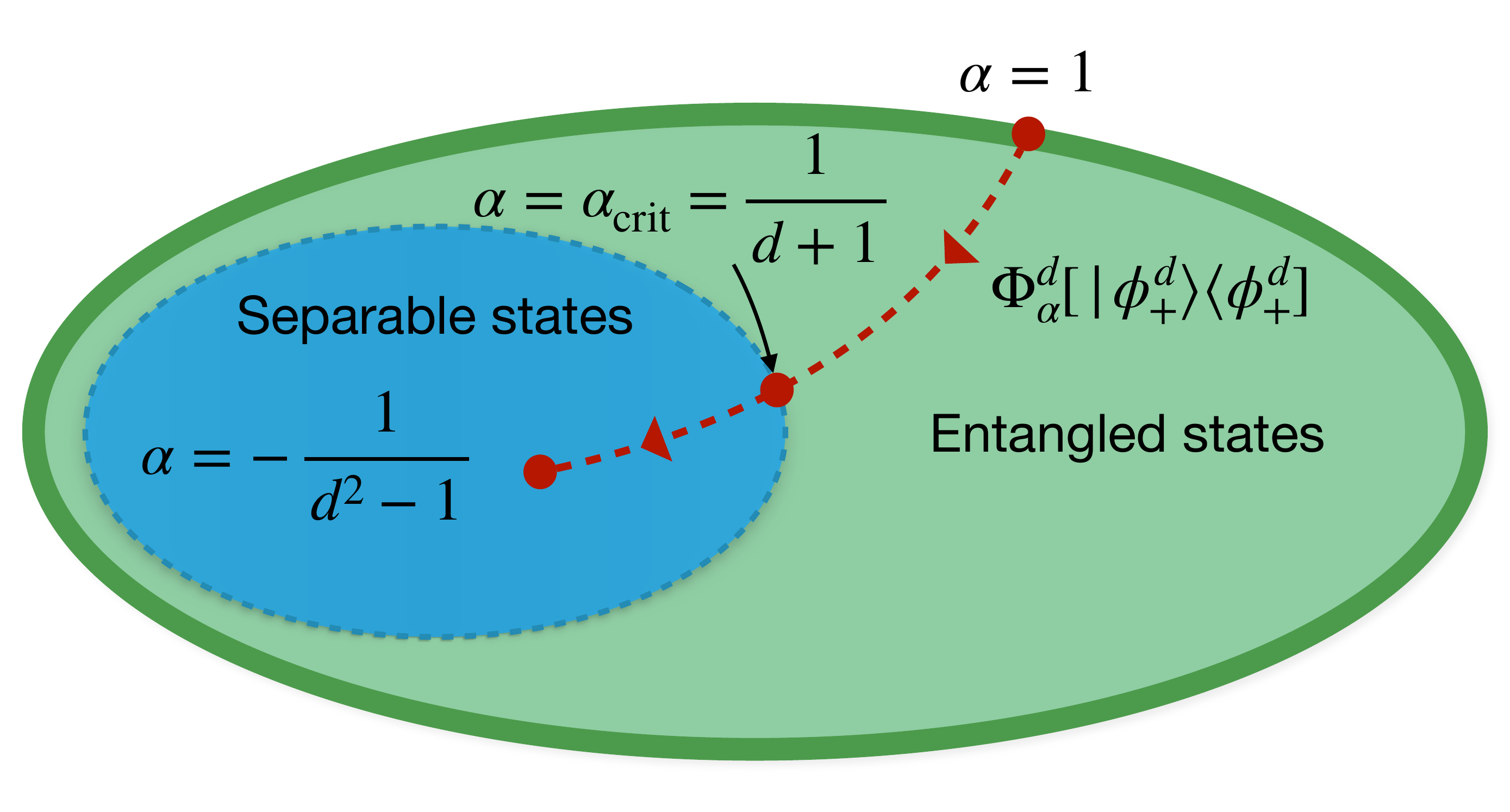}
\caption{Applying the confusion scheme to the case of the maximally entangled stated depolarizing.
}
\label{fig:DepIntuition}
\end{figure}

Consider the state $\sigma_{\alpha}^d:=\Phi_\alpha^d[\ket{\phi_+^d}\bra{\phi_+^d}]$ resulting from the action of the depolarizing channel on the maximally entangled state:
\begin{equation}
    \ket{\phi_+^d} = \frac{1}{\sqrt{d}} \sum_{i=0}^{d-1} \ket{i}\otimes\ket{i},
\end{equation}
where $\{\ket{i}\}_{i=0}^{d-1}$ denote computational basis states.
According to the depolarization channel properties, $\sigma_{\alpha}^d$ is entangled for $(d+1)^{-1} < \alpha \leq 1$, and 
separable for $-(d^2-1)^{-1} \leq \alpha \leq (d+1)^{-1}$.
At the same time, it is known that $\Phi_{\alpha}^d[\cdot]$ is entanglement breaking if and only if $-(d^2-1)^{-1} \leq \alpha \leq (d+1)^{-1}$~\cite{Horodetski1999,Moravcikova2010}.
Thus, the critical `phase transition' point for the depolarizing channel is given by $\alpha_{\rm crit} = (d+1)^{-1}$ (see Fig.~\ref{fig:DepIntuition}).

\begin{figure*}[]
\includegraphics[width=1\textwidth]{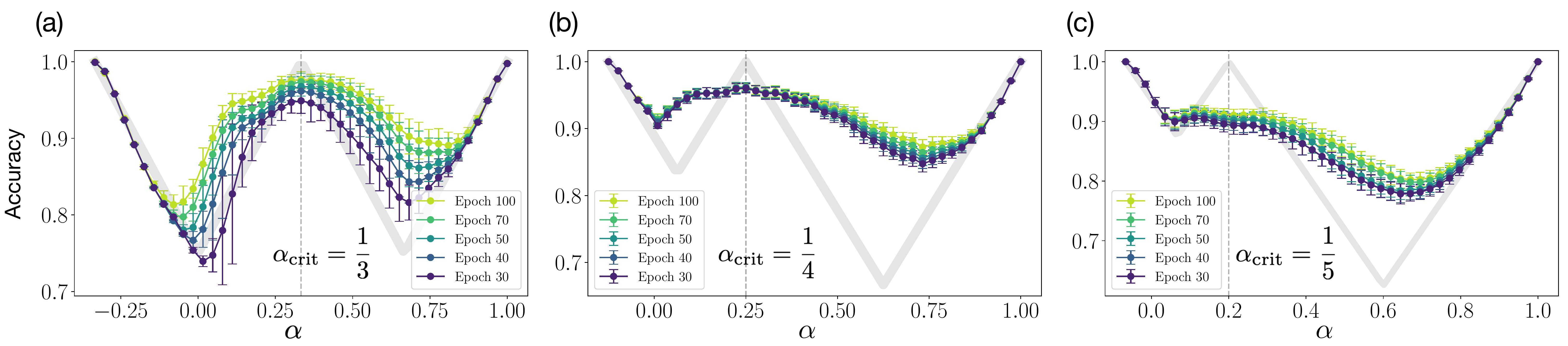}
\caption{W shapes of the FFNN accuracy for the case of depolarizing two-qubit (a), two-qutrit (b), and two-ququart (c) maximally entangled states.
The ideal expected W shapes are shown with thick grey lines. 
Vertical dashed line indicate the expected critical points.}
\label{fig:Depolarization-Wshapes}
\end{figure*}

To construct the dataset for the confusion scheme, we take a sequence of $N$ states
\begin{equation}
	\rho_i=\left(U_A^i\otimes U_B^i\right) \sigma_\alpha^d\left(U_A^i\otimes U_B^i\right),
\end{equation}
where $i=0,1,\ldots,N-1$, depolarizing parameter is evenly distributed over the corresponding domain:
\begin{equation}
	\alpha_i=-\frac{1}{d^2-1}+\frac{i}{N-1}\left(1+\frac{1}{d^2-1}\right),
\end{equation}
and $U_{A(B)}^i$ are random local $d\times d$ unitary operators distributed uniformly according to the Haar measure~\cite{Haar2004} (see details of the generation of $U_{A(B)}$ in Appendix~\ref{apx:Haar}).
We consider two-qubit, two-qudtrit, and two-quaqart cases by fixing $d=2,3,4$ respectively.
The additional randomization procedure with local unitaries provides the correct performance of the confusion scheme: 
we force the neural network to extract features related to correlation between $A$ and $B$ that are
insensitive to these local transformations.
The resulting states $\rho_i$ are then transformed to probability vectors $\vec{p}_i$ according to the procedure described in Sec.~\ref{sec:dataprep}.

The resulting W shapes of accuracy functions, obtained by estimating a FFNN performance for different epochs of training, are shown in Fig.~\ref{fig:Depolarization-Wshapes}.
The presented curves are obtained by averaging W shapes obtained by training a neural network on $100$ independently sampled datasets of length $N=1400$.
Among each set, $0.7N$ and $0.3N$ entries are used for training and testing correspondingly.
The configuration of the employed FFNN is described in Appendix ~\ref{apx:NN}.
As an estimate of the neural network performance, we choose the maximum value of the classification accuracy achieved during the training process on the test data set. 
This corresponds to the early stopping of the learning process and allows us to avoid the performance underestimation. One can capture the universal W shapes properly  separating the subspaces of entangled and separable quantum states in Fig.~\ref{fig:Depolarization-Wshapes}.

We also note that the confusion learning scheme results can be obtained for relatively small dataset sizes ($200$-$1500$ samples).

\subsection{Entanglement breakdown phase diagrams}\label{sec:PhaseDiagrams}

Our next step is to study entanglement breakdown for two-parametric quantum channels.
It allows us to investigate a potential of the confusion scheme to reveal whole entanglement breakdown `phase diagrams' in the parameter space of quantum channels, rather than a single critical point of the entanglement breakdown.

First, let us consider a channel $\widetilde{\Phi}_{\alpha_1,\alpha_2}$ obtained as two qubit depolarizing channels $\Phi^2_{\alpha_1}$ and $\Phi^2_{\alpha_2}$  acting on distinct qubits of a two-qubit state $\rho$:
\begin{multline}\label{eq:SepDepChannel}
	\widetilde{\Phi}_{\alpha_1,\alpha_2}[\rho]=(\Phi^2_{\alpha_1} \otimes \Phi^2_{\alpha_2} )[\rho] 
	\\ =\alpha_{1} \alpha_{2} \rho +
	\frac{1}{2}(1-\alpha_{1})\alpha_{2}\mathbf{1}\otimes \rho_{B} + \\
	 +\frac{1}{2}\alpha_{1}(1-\alpha_{2})\rho_{A}\otimes \mathbf{1} + \frac{1}{4}(1-\alpha_{1})(1-\alpha_{2})\mathbf{1}\otimes\mathbf{1}.
\end{multline}
Here $\rho_{A}$ and $\rho_{B}$ are reduced qubit states of $\rho$, and $\alpha_{1}, \alpha_{2}$ are depolarizing parameters.

\begin{figure*}[]
\includegraphics[width=1.\textwidth]{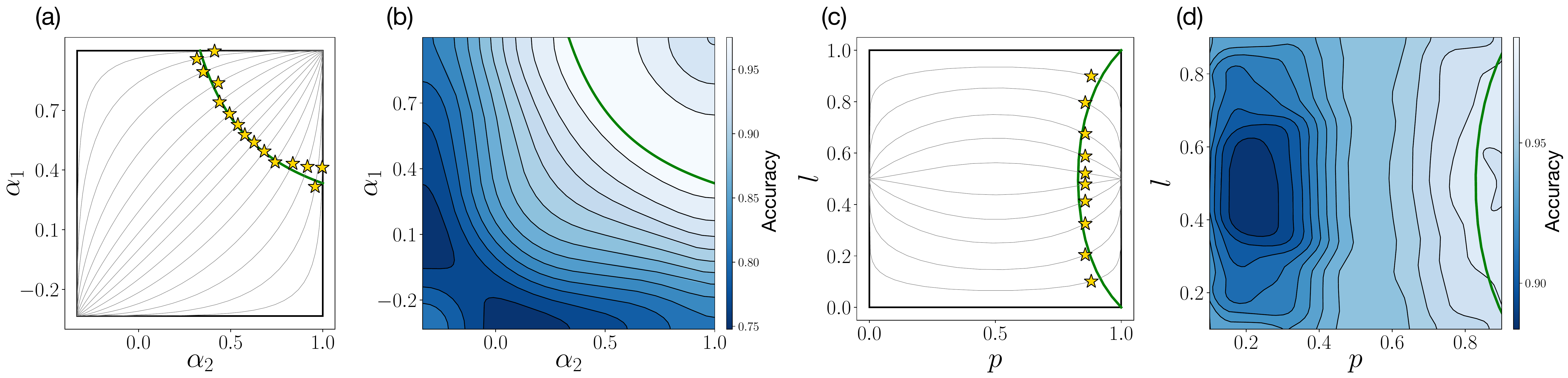}
\caption{Results of applying the confusion scheme for two-parametric channels $\widetilde\Phi_{\alpha_1,\alpha_2}$ in (a), (b) and $\widetilde{\cal A}_{l,p}[\cdot]$ in (c), (d) are shown. 
In (a) and (c) curves in parametric space, used within the confusion scheme, and obtained critical points are shown. 
In (b) and (d) the obtained heat maps of W shapes on studied curves are presented. Bold green lines show rigorous theoretical results on the entanglement-breaking regions.
All the results correspond to 100 epochs of the training process.}
\label{fig:Phase-Diagrams}
\end{figure*}

Since the confusion scheme can detect transitions between entangled and separable states for one-parameter dataset only, to apply the scheme to the multi-parameter channel we construct a family of trajectories in the channel's parameter space 
$(\alpha_1,\alpha_2)$ [see Fig.~\ref{fig:Phase-Diagrams}(a,b)].
Each trajectory is the parametric family of quantum states obtained by first applying $\widetilde{\Phi}_{\alpha_1,\alpha_2}$ to the maximally entangled state $\ket{\phi_+^2}$ for $(\alpha_1, \alpha_2)$ belonging to some curve, 
and then applying a pair of random local unitaries. 
We use curves starting from the point of $\alpha_1=\alpha_2=1$ and ending at  $\alpha_1=\alpha_2=-1/3$.
This choice allows us to connect the 
the maximally entangled state with the maximally mixed one.
The curves covering the whole space shown in Fig.~\ref{fig:Phase-Diagrams} (see also Appendix~\ref{apx:Param}).

The obtained phase transition points and corresponding heat map of obtained W shapes are shown in Fig.~\ref{fig:Phase-Diagrams}(a) and (b) correspondingly.
In Fig.~\ref{fig:Phase-Diagrams}(a,b) we also show an exact solution for the bound of the entanglement annihilating region shown in Ref.~\cite{Filippov2012}.
One can see a quite good correspondence between results confusion scheme and rigorous theoretical results, and conclude that the confusion scheme is suitable for studying entanglement breaking transitions. 

Essentially, this result is not exclusive for the depolarizing channel and can be obtained for other channels. 
We also consider a generalized single-qubit amplitude damping channel $\mathcal{A}_{l,p}$ given by corresponding Kraus operators
\begin{equation}
    \begin{aligned}
	    A_{l,p}^{(1)}&=\sqrt{l}
	    \begin{bmatrix}
	1&0\\
	0& \sqrt{1-p}\\
        \end{bmatrix},~       
        &A_{l,p}^{(2)}&=\sqrt{l}
	\begin{bmatrix}
        0& \sqrt{p}\\
        0&0\\
        \end{bmatrix},      \\   
        A_{l,p}^{(3)}&=\sqrt{1-l}
	\begin{bmatrix}
        \sqrt{1-p}&0 \\
        0&1\\
        \end{bmatrix},~     
         &A_{l,p}^{(4)}&=\sqrt{1-l}
	\begin{bmatrix}
        0&0 \\
        \sqrt{p}&0\\
        \end{bmatrix},        
    \end{aligned}
\end{equation}
where $l \in[0,1]$, $p\in[0,1]$. 
We consider an action of the generalized single-qubit amplitude damping channel on single qubit of a two-qubit state, so the resulting two-qubit channel reads
\begin{multline} \label{eq:GenAmpDamp}
	\widetilde{\mathcal{A}}_{l,p}[\rho]=(\mathcal{A}_{l,p} \otimes {\rm Id})[\rho] \\= \sum_{i=1}^{4}\left(A^{(i)}_{l,p}\otimes {\bf 1}\right)\rho \left(A^{(i)}_{l,p}\otimes {\bf 1}\right)^{\dagger},
\end{multline}
where ${\rm Id}$ is the identity channel.

Similarly, for the case of channel~\eqref{eq:GenAmpDamp}, we reconstruct the phase diagram by preparing a dataset based on states $\widetilde{\mathcal{A}}_{l,p}[\ket{\phi^2_+}\bra{\phi^2_+}]$, randomized by local unitary operators.
The values of parameters $l$ and $p$ are taken from the curves connecting points $l=1/2$, $p=0$ (the maximally entangled state) as well as $l=1/2$ and $p=1$ (the maximally mixed state).
The precise curves parametrization is given in Appendix~\ref{apx:Param}.

The results on applying the confusion scheme are shown in Fig.~\ref{fig:Phase-Diagrams}~(c,d). 
As in the case of depolarizing channels, the comparison with the theoretical results related to the entanglement breaking region~\cite{Filippov2012} shows that the `learning by confusion' method gives quite accurate position of the transition.

We see that by using the confusion scheme, it is possible to reconstruct the phase diagram for the case of a depolarizing channel and a generalized amplitude damping channel. 
This proposes that the confusion scheme allows us to explore the entanglement-related properties of other quantum channels.

\section{Entanglement classification}\label{sec:EntClassification}

\begin{figure}[h!]
\includegraphics[width=0.9\linewidth]{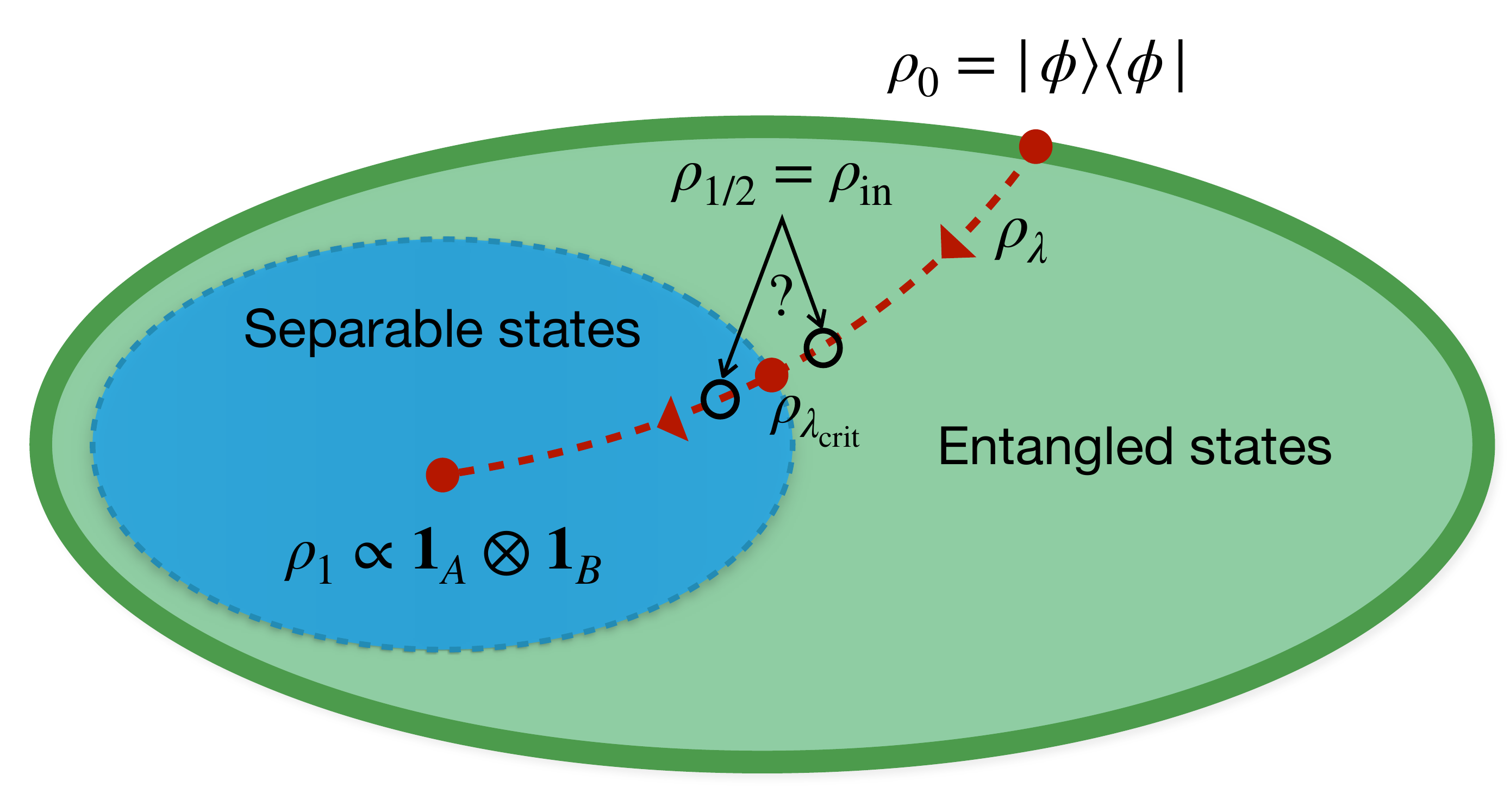}
\caption{
The use of the confusion scheme for determining whether the given state $\rho_{\rm in}$ is entangled or not.
We design a parametric state $\rho_\lambda$, such that $\rho_0$ is maximally entangled, $\rho_1$ us maximally mixed, and $\rho_{1/2}=\rho_{\rm in}$.
Then we run the confusion scheme to the trajectory of states $\rho_\lambda$ to find a critical point $\lambda_{\rm crit}$.
Finally, by comparing $\lambda_{\rm crit}$ with 1/2 we conclude whether $\rho_{\rm in}$ is entangled or not.}
\label{fig:class}
\end{figure}

Here we extend the approach for revealing entanglement breakdown using the confusion scheme to the problem of recognizing whether a given state is entangled or not.
As discussed above, the confusion learning scheme is capable of detecting transitions between entangled and separable states as a result of some decoherence processes. 
At the same time, finding the parameter corresponding to the transition point makes it possible to classify the entanglement of states that make up the corresponding dataset. 
Thus, in order to classify the entanglement of arbitrary quantum state $\rho_{\rm in}$ we can include it in a specially designed dataset corresponding to transformation of some maximally entangled state to some separable state (see Fig.~\ref{fig:class}). 
We consider a parametric state in form of thermal state:
\begin{equation}
    \begin{aligned}
	    &\rho_{\lambda} = \frac{1}{Z}  e^{- \beta (\lambda) H(\lambda)},
	    &Z = {\rm Tr}[e^{- \beta (\lambda) H(\lambda)}],
	\end{aligned}
\end{equation}
satisfying the following conditions:
\begin{equation} \label{eq:conditions-on-beta}
    \rho_0=\ket{\phi}\bra{\phi},
    \quad 
    \rho_{1/2}=\rho_{\rm in}, 
    \quad
    \rho_{1}=d_A^{-1}d_B^{-1}{\bf 1}_A\otimes {\bf 1}_B,
\end{equation}
where $\ket{\phi}\bra{\phi}$ is the maximally entangled state. For detailed description of the parametrization see Appendix \ref{apx:Class}.

\begin{figure*}[]
\includegraphics[width=1.\textwidth]{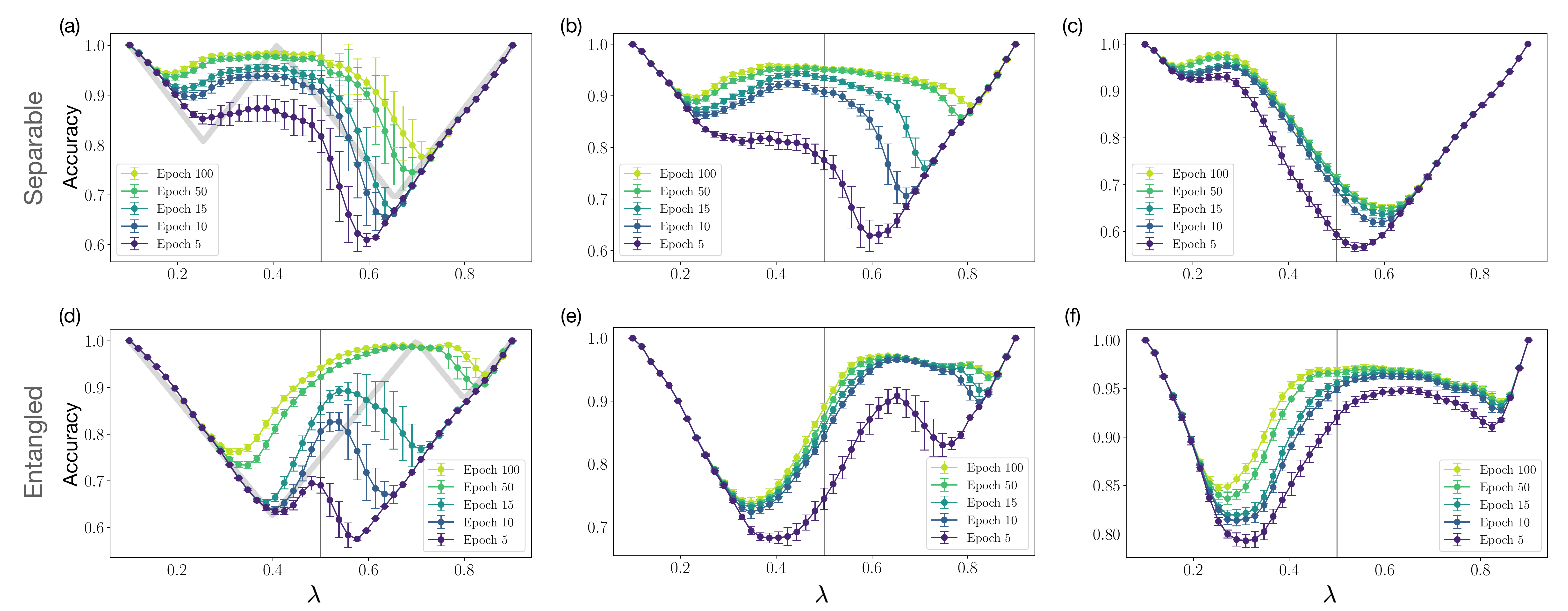}
\caption{Results of applying the confusion-based entanglement classification procedure for of separable states $\Phi_{0.2}^2(\rho_{+}^{2})$ (a), $\Phi_{0.2}^3(\rho_{+}^{3})$ (b),
$\rho_{\rm cq}$ (c), and
entangled states
$\Phi_{0.9}^2(\rho_{+}^{2})$ (d),
$\Phi_{0.9}^3(\rho_{+}^{2})$ (e),
$\sigma_{0.2}$ (f).
The regions on the left and right of the W shapes middle peaks correspond to entangled and separable states respectively, while the analyzed states correspond to $\lambda=1/2$.
Gray bold color in (a) and (d) shows ideal W shape computed using known necessary and sufficient conditions for entanglement in two-qubit states.}
\label{fig:ClassificationResults}
\end{figure*}

The dataset for the confusion scheme is generated by taking values $\lambda_i$ evenly distributed over $[0.1,0.9]$, and calculating states $\rho_{\lambda_i}$ additionally randomized by local operations.
Then the critical point $\lambda_{\rm crit}$, obtained from the confusion scheme, is compared with 1/2.
The state is supposed to be entangled $\lambda_{\rm crit}>1/2$, and it is separable otherwise. 

If Fig.~\ref{fig:ClassificationResults} we show the resulting W shapes of accuracy function for various entangled (a-c), and separable (d-f) states.
Namely, in Fig.~\ref{fig:ClassificationResults}(a,d) and Fig.~\ref{fig:ClassificationResults}(b,e), the states obtained using depolarizing channel acting on two-qubit and two-qutrit maximally entangled states are considered.
In Fig.~\ref{fig:ClassificationResults}(c) the results for a two-qutrit `classical-quantum' state
\begin{equation}
    \begin{aligned}
        \rho_{\rm cq} &= \sum_{i=0}^2 \ket{i}\bra{i} \otimes \ket{\chi_{i}}\bra{\chi_{i}},\\
        \ket{\chi_0} &= 3^{-1/2}(\ket{0}+\ket{1}-\ket{2}),\\
        \ket{\chi_1} &= 3^{-1/2}(\ket{0}-\ket{1}+\ket{2}),\\
        \ket{\chi_2} &= 3^{-1/2}(-\ket{0}+\ket{1}+\ket{2})\\
    \end{aligned}
\end{equation}
are shown.
In Fig.~\ref{fig:ClassificationResults}(f) we analyze an entangled two-qutrit state $\sigma_{0.2}$, where 
\begin{equation}
\sigma_{a} = \frac{1}{8a + 1}\begin{bmatrix} 
						    a & 0 & 0 & 0 & a & 0 & 0 & 0 & a\\
						    0 & a & 0 & 0 & 0 & 0 & 0 & 0 & 0\\
						    0 & 0 & a & 0 & 0 & 0 & 0 & 0 & 0\\
						    0 & 0 & 0 & a & 0 & 0 & 0 & 0 & 0\\
						    a & 0 & 0 & 0 & a & 0 & 0 & 0 & a\\
						    0 & 0 & 0 & 0 & 0 & a & 0 & 0 & 0\\
						    0 & 0 & 0 & 0 & 0 & 0 & \frac{1 + a}{2} & 0 & a\\
						    0 & 0 & 0 & 0 & 0 & 0 & 0 & a & \frac{\sqrt{1 - a^2}}{2}\\
						    a & 0 & 0 & 0 & a & 0 & \frac{\sqrt{1 - a^2}}{2} & 0 & \frac{1 - a}{2}\\
					\end{bmatrix},
\end{equation}
which is special in that it has a positive partial transpose (PPT)~\cite{Horodecki1997}.

We see that for all considered states the developed confusion-based entanglement classification provides the correct answer.
Notably, the scheme still works correctly even in the case where commonly used entanglement measure as negativity fails to recognize a state as entangled.
In this way, the confusion scheme demonstrates capabilities of determining whether the arbitrary given state is entangled or not.

\begin{figure}[h!]
\includegraphics[width=\linewidth]{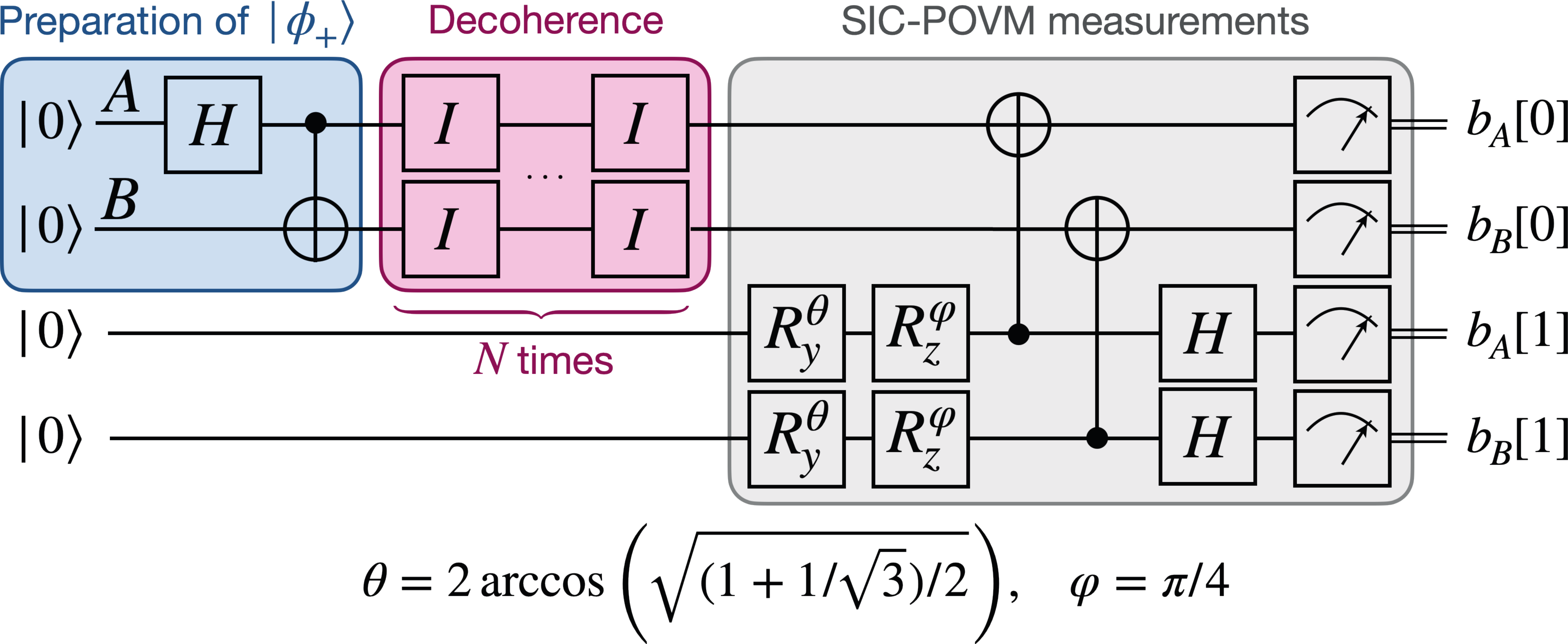}
\caption{Quantum circuit that is used for experimental study of an entanglement decay with IBM quantum processor.
Standard notations of Hadamard, controlled-NOT, and Pauli rotation gates are used.
Pairs of single-bit values $(b_A[0],b_A[1])$ and $(b_B[0],b_B[1])$ encode results of SIC-POVM measurements of qubits $A$ and $B$, correspondingly.}
\label{fig:circ}
\end{figure}

\begin{figure*}[]
\includegraphics[width=1.\textwidth]{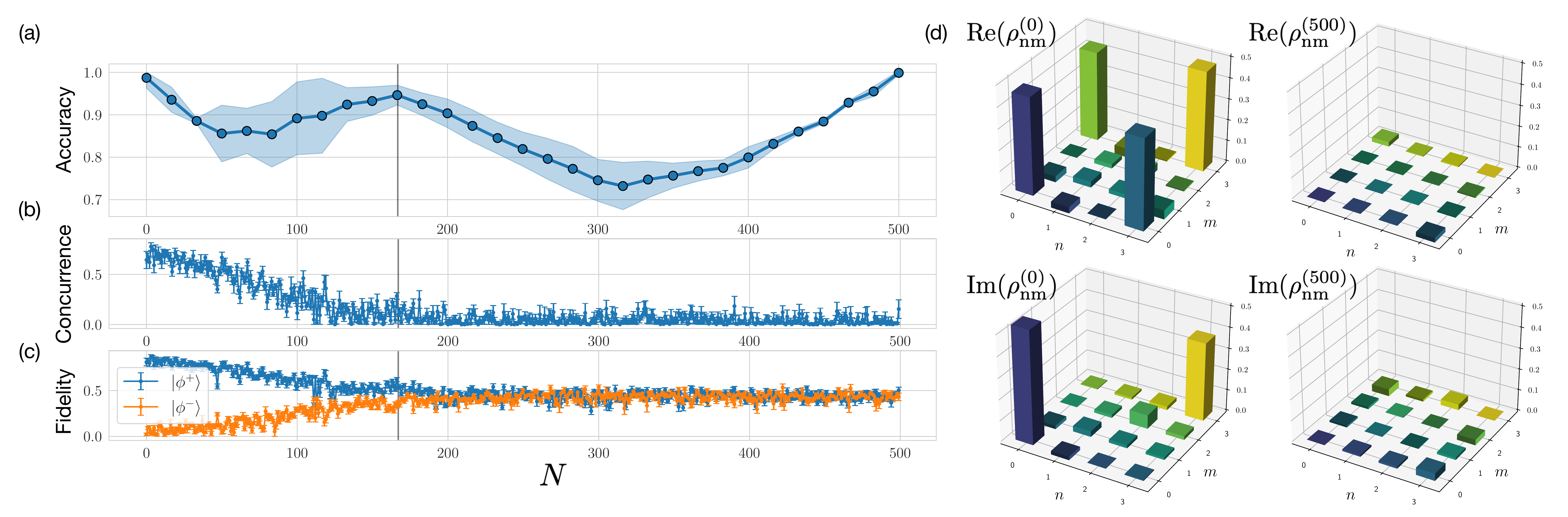}
\caption{
    The experimental results obtained with IBM quantum processor.
    In (a) the results of averaging accuracy function of generated datasets is shown. 
    Bold dots correspond to mean values, highlighted region show a spread between the minimal and the maximal values.
    In (b) the concurrence of $\rho^{(N)}$ is depicted.
    In (c) the fidelities of $\rho^{(N)}$ with respect to Bell states $\ket{\phi_+}$ and $\ket{\phi_-}$ are presented.
    In (d) the reconstructed density matrices of $\rho^{(0)}$ and $\rho^{(500)}$, which corresponds to the extreme points of the confusion scheme dataset, are shown.}
\label{fig:IBMQ}
\end{figure*}

\section{Detecting entanglement with A quantum processor}\label{sec:IBM}

In this section, we consider an application of the confusion scheme for studying an entanglement-separability transition appearing due to decoherence process in modern noisy intermediate scale quantum (NISQ) devices.
Namely, we consider a decay of entanglement in a state prepared using real $5$-qubits superconducting processor IBM Athens.
For this purpose we run a 4-qubit circuit shown in Fig.~\ref{fig:circ}.
The circuit consists of three main parts: (i) preparing the maximally entangled Bell state $\ket{\phi_+} \equiv  \ket{\phi_+^2}$ on the first two qubits, (ii) applying a decoherence channel to the prepared entangled state by using a sequence of $N$ identity operators, and (iii) performing SIC-POVMs measurements of the first two qubits by employing two additional qubits. 
We vary the integer parameter $N$ from 0 to 500, and use it as a confusing parameter.
For $N=0$ the state is expected to be close to maximally entangled $\ket{\phi_+}$, and for $N=500$ the state is expected to be separable due to decoherence.

To obtain a probability (frequency) distribution that is used for making up a confusion scheme dataset we run the designed circuit $N_{\rm shots}=8192$ times for each value of $N=0,\ldots,500$.
We also reconstruct a two-qubit density matrix $\rho^{(N)}$ with a standard quantum state tomography approach.
Obtained density matrices are used to `bootstrap' additional confusion scheme datasets (in result, 30 complete datasets are considered).

To verify the results of the confusion scheme, for each $\rho^{(N)}$ we compute the value of concurrence ${\cal C}(\rho^{(N)})$ defined by the following expression~\cite{Wootters1997,Wootters1998}:
\begin{equation}
	\mathcal{C}(\rho) = \max\{0, \nu_1 - \nu_2 - \nu_3 - \nu_4\},
\end{equation}
where $\nu_1,\ldots,\nu_4$ are the eigenvalues in decreasing order of the Hermitian matrix
\begin{equation}
	R = \sqrt{\sqrt{\rho} (\sigma_y \otimes \sigma_y) \rho^* (\sigma_y \otimes \sigma_y) \sqrt{\rho}},
\end{equation}
$\sigma_y$ is the standard Pauli matrix, and $\rho^*$ denotes a complex conjugate of $\rho$.
Remind that ${\cal C}(\rho)>0$ is the necessary and sufficient condition for $\rho$ to be entangled.
To provide a better intuition of the realized physical process, we also compute fidelities of $\rho^{(N)}$ with respect to $\ket{\phi_+}$ and $\ket{\phi_-}= 2^{-1/2} (\ket{00}-\ket{11})$.

The obtained results are shown Fig.~\ref{fig:IBMQ}.
In (a) we presented the results of averaging accuracy functions for generated datasets.
We see a clear W shape of an average accuracy with a middle peak at $N=165$.
As it is shown in Fig.~\ref{fig:IBMQ}(b), this value of $N$ almost exactly matches with point, where the concurrence drops to 0.
This point also corresponds to turning of the initial state $\ket{\phi_+}$ into approximately even mixture of $\ket{\phi_+}\bra{\phi_+}$ and $\ket{\phi_-}\bra{\phi_-}$ [see Fig.~\ref{fig:IBMQ}(d,e)].

We can conclude the confusion scheme properly reveals an `entanglement-breakdown transition' in experimental data.
In this way, the confusion scheme demonstrates a potential to be used for studying properties of modern NISQ devises.

\section{Conclusion and outlook}\label{sec:Conclusion}

In this work, we have proposed a machine-learning-based approach for the analysis of the entanglement breakdown for parametrized families of quantum states.
The used heuristic `learning by confusion' correctly predicts the critical value, at which the transition from entangled to separable classes of states within the family takes place, without analyzing the structure of such states. 
We have applied the proposed method to the cases of two-qubit, two-qutrit and two-ququart quantum states and obtain corresponding W shapes. 
We have obtained W shapes that show the correct transition between the separable state and entangled state for each case. 
We have demonstrated the possibility of reconstructing the phase diagram for the case of a local depolarizing channel and for the case of a generalized amplitude damping channel. 
This allows us to get the most complete picture of the state under consideration and to understand in which region we observe entangled states, and which region is areas of entanglement-breakdown. This makes the proposed scheme suitable for studying the properties entanglement-breaking and entanglement-annihilating properties, e.g. $N$-local entanglement annihilating ($N$-LEA) property\cite{Filippov2012}, of various quantum channels.
We have demonstrated that the `learning by confusion' scheme can be successfully extended to the case of arbitrary quantum states and allows to classify a given bipartite state as an entangled or a separable one. 

Also in this work, we have conducted an experiment using the IBM Athens quantum processor via cloud platform. 
The problem of detecting entanglement of quantum states is important and relevant for many experimental implementations in the quantum technologies field, for example, quantum communications and quantum computing. 
As a result of the experiment, where the maximally entangled two-qubit state was initially prepared, 
we have shown that using the data obtained as a result of performing quantum circuit, the transition 
between the entangled and separable states can be detected. 
This important result shows that the learning by confusion scheme can be applied to real experimental data obtained 
with existing NISQ devices and can give us an answer to the question of where the transition between entangled and separable quantum states lies.

One of the shortcomings of the proposed approach is the fact that we have to use the result of SIC-POVM measurements, which are hard to obtain for large enough quantum systems. Therefore, an important question is the sensitivity of this approach to an incompleteness in a quantum state characterization. Also, the development of our approach can be the use of the method described in the work \cite{Liu2018} for study the two-parameter quantum channels.
This is the subject of a subsequent study. 

\section*{Acknowledgments}
We acknowledge use of the IBM Q Experience for this work. 
The views expressed are those of the authors and do not reflect the official policy or position of IBM or the IBM Q Experience team.
The authors thank E. Tiunov and A. Ulanov for fruitful discussions.
The work is supported by the Russian Science Foundation (project 20-42-05002; the development of the method in Sec.~\ref{sec:W}) 
and UMNIK grant (Agreement 16576GU/202 form 02.06.2020; development of the neural-network based software in Sec.~\ref{sec:W}).
We also acknowledge the support from Leading Research Center on Quantum Computing (Agreement 014/20; generalizaiton of the method in Sec.~\ref{sec:EntClassification} and application to real quantum computers in Sec.~\ref{sec:IBM}).

\appendix
 
\section{Geterating SICs with Weyl groups}\label{apx:SICPOVM}

Each SIC-POVM effect can be constructed by starting with a fiducial vector, and acting upon it with the elements of some group. 
In all these known cases (but one) the group that generates a SIC-POVM is a Weyl-Heisenberg group~\cite{Stacey2017,Stacey2017-2,Waldron2018}. 

The Weyl-Heisenberg group is defined as follows. 
Let us fix the value of the dimension of the system $d$.
The shift $X$ and phase $Z$ operators have the form
\begin{equation}
    X \ket{j}=\ket{j+1}, \quad Z\ket{j}=\omega^i\ket{j},
\end{equation}
where $\omega:=e^{2\pi \imath/d}$.
Then the Weyl-Heisenberg displacement operator is
\begin{equation}
    D_{n} = (-e^{\imath\pi/d})^{n_1n_2}X^{n_1}Z^{n_2},
\end{equation}
where $n=(n_1,n_2)$ is a multi-index, with $n_i\in \mathbb{Z}_d$.

A vector $\ket{\psi}$ is called fiducial if it satisfies the following conditions:
\begin{equation}
    |\bra{\psi}D_{n}\ket{\psi}|^2=\frac{d\delta_{n,0}+1}{d+1}, \quad \langle{\psi}|\psi\rangle=1.
\end{equation}
Then a SIC-POVM can be constructed as the orbit of a fiducial vector $\ket{\psi}$ under the action of the Weyl-Heisenberg displacement operators:
\begin{equation}
    E_{n}= \frac{1}{d} D_n\ket{\psi}\bra{\psi}D_n^{\dagger}.
\end{equation}

Fiducial vectors for cases of $d=2,3,4$, which are considered in the main text, are as follows:
\begin{equation}
    \begin{aligned}
        \ket{\psi}&=\frac{1}{\sqrt{6}}
        \begin{bmatrix}
            \sqrt{3+\sqrt{3}} \\
            e^{i\pi/4}\sqrt{3-\sqrt{3}} 
        \end{bmatrix},\\
        \ket{\psi}&=\frac{1}{\sqrt{2}}
            \begin{bmatrix}
            0\\
            1 \\
            -1
            \end{bmatrix},\\
        \ket{\psi} &= \frac{1}{\sqrt{5 + \sqrt{5}}}
         \begin{bmatrix} 
		 \sqrt{2 + \sqrt{5}}\\
		 1\\
		 1\\
		 1
	\end{bmatrix}.
    \end{aligned}
\end{equation}

\section{Random unitary generation}\label{apx:Haar}
For generation of the random $N\times N$ unitary matrices we use the algorithm described in Ref.~\cite{Mezzadri2006}. 
The algorithm consists of the following steps.
\begin{enumerate}
    \item Generate matrix ${\bf Z} = {\bf X} + \imath {\bf Y}$, where ${\bf X}, {\bf Y}$ are $N \times N$ matrices with random entries that are normally distributed with zero mean and unit variance.
    \item Compute QR-decomposition ${\bf Z} = {\bf Q} {\bf R}$.
    \item Compute diagonal matrix $\Lambda$ with $\Lambda_{i,i} = {\bf R}_{i,i} / |{\bf R}_{i,i}|$;
    \item Compute ${\bf U} = {\bf Q} \Lambda$ which is uniformly distributed under Haar measure.
\end{enumerate}

\section{Details on the neural network}\label{apx:NN}

In this work, we use an Adam optimizer to train the neural network with the set of parameters and network configuration listed in Tables~\ref{tbl:FFNN1}-\ref{tbl:FFNN4}.
\begin{table*}[h!]
	\caption{Structure and parameters of the FFNN for numerical experiments in Sec.~\ref{sec:W}.}
	
		\begin{tabular}{p{0.33\linewidth}|p{0.1\linewidth}|p{0.1\linewidth}|p{0.1\linewidth}}
			 & $2\otimes 2$ case & $3\otimes 3$ case &  $4 \otimes 4$\\ \hline \hline
			Configuration of the network \\ \hline \hline
			Number of neurons at the input layer & $16$ &  $81$  & $256$\\ \hline
			Number of neurons at the hidden layer & $32$ &  $162$ & $512$\\ \hline
			Number of neurons at the output layer& $2$ & $2$ &  $2$ \\ \hline  \hline  
			Set of parameters  \\ \hline \hline
			Learning rate, $\rm{lr}$   & $10^{-3}$	 &	$10^{-3}$ & $10^{-5}$	  \\
			Weight decay, $\varepsilon$	& $10^{-4}$	 &	$10^{-5}$ &	 $10^{-5}$ \\
			
			Data size	& $1400, 1400$	 &	$1400$ & $1400$	\\
			Batch size	& $50$	 &	$50$	 &  $50$\\		
		\end{tabular}
		\label{tbl:FFNN1}
	\end{table*}

\begin{table*}[h!]
\centering
\caption{Structure and parameters of the FFNN for numerical experiments in Sec.~\ref{sec:PhaseDiagrams}.}
\begin{tabular}{p{0.4\linewidth}|p{0.1\linewidth}}
			Configuration of the network \\ \hline \hline
			Number of neurons at the input layer & $16$ \\ \hline
			Number of neurons at the hidden layer & $64$ \\ \hline
			Number of neurons at the output layer & $2$ \\ \hline  \hline  
			Set of parameters   \\ \hline \hline
			Learning rate, $\rm{lr}$   & $2\times10^{-3}$  \\
			Weight decay, $l_1$	& $2\times 10^{-5}$	\\
			Data size  & $1000$	 \\
			Batch size & $50$	 \\		
\end{tabular}
\label{tbl:FFNN2}
\end{table*}

\begin{table*}[h!]
\caption{Structure and parameters of the FFNN for numerical experiments in Sec.~\ref{sec:EntClassification}.}

	\begin{tabular}{p{0.33\linewidth}|p{0.1\linewidth}|p{0.1\linewidth}}
		 & $2\otimes 2$ case & $3\otimes 3$ case \\ \hline \hline
		Configuration of the network \\ \hline \hline
		Number of neurons at the input layer & $16$ &  $81$ \\ \hline
		Number of neurons at the hidden layer & $32$ &  $162$ \\ \hline
		Number of neurons at the output layer& $2$ & $2$ \\ \hline  \hline  
		Set of parameters  \\ \hline \hline
		Learning rate, $\rm{lr}$   & $10^{-3}$	 &	$10^{-2}$ 	  \\
		Weight decay, $\varepsilon$	& $10^{-3}$	 &	$10^{-4}$  \\
		
		Data size	& $1400$	 &	$1400$ 	\\
		Batch size	& $100$	 &	$100$	 \\		
	\end{tabular}
	\label{tbl:FFNN3}
\end{table*}

\begin{table*}[h!]
\centering
\caption{Structure and parameters of the FFNN for numerical experiments in Sec.~\ref{sec:IBM}.}
\begin{tabular}{p{0.4\linewidth}|p{0.1\linewidth}}
			Configuration of the network \\ \hline \hline
			Number of neurons at the input layer & $16$ \\ \hline
			Number of neurons at the hidden layer & $16$ \\ \hline
			Number of neurons at the output layer & $2$ \\ \hline  \hline  
			Set of parameters   \\ \hline \hline
			Learning rate, $\rm{lr}$   & $2\times10^{-4}$  \\
			Weight decay, $l_1$	& $2\times 10^{-3}$	\\
			Data size  & $500$	 \\
			Batch size & $50$	 \\		
\end{tabular}
\label{tbl:FFNN4}
\end{table*}

\section{Curves parameterization}\label{apx:Param}

For reconstructing phase diagram for the local two-qubit depolarizing channels, the curves are given by the parametric equations with parameter $t$:
\begin{equation}
\begin{aligned}
        \begin{cases}
        \alpha_1(t) = \frac{4}{3}\cos(t)^{\frac{2}{n}} - \frac{1}{3},\\
        \alpha_2(t), = -\frac{4}{3}\sin(t)^{\frac{2}{n}} + 1
        \end{cases} ~~~\text{for}~~n \leq 1;
        \\
         \begin{cases}
        \alpha_1(t) = -\frac{4}{3}\cos(t)^{\frac{2}{n - 1}} + 1,\\
        \alpha_2(t) = \frac{4}{3}\sin(t)^{\frac{2}{n - 1}} - \frac{1}{3},
        \end{cases} \text{for}~~n > 1,
	 \end{aligned}
\end{equation}
where  $n \in \{0.3, 0.4, \ldots, 1, 1.3, 1.4, \ldots, 1.9 \}$ and $t \in[0, \frac{\pi}{2}]$.

For the case of reconstructing phase diagrams for the generalized amplitude damping channel, the curves parameterization equations have the following form:
\begin{equation}
\begin{aligned}
	\begin{cases}\label{damping-parametrization}
		\beta = \frac{1}{2} p^{|t|} (1 - p)^{|t|} + \frac{1}{2} ~~ t \geq 0\\
		\beta = -\frac{1}{2} p^{|t|} (1 - p)^{|t|} + \frac{1}{2} ~~ t < 0,\\
	\end{cases}
\end{aligned}
\end{equation}
where $t\in \{ \pm 0.1, \pm 0.25, \pm 0.5, \pm 0.83, \pm 1.5\}$ and $p \in (0, 1)$.

\section{Entanglement classification scheme}\label{apx:Class}
In order to construct a dataset, let us consider an input density matrix $\rho_{\rm in}$ as a thermal state of unit temperature:
\begin{equation}\label{eq:thermal-state}
    \rho_{\rm in} = \exp(- H_{\rm in}),
\end{equation}
where $H_{\rm in}= \ln\rho_{\rm in}$ is the corresponding `effective Hamiltonian'. 
In what follows, we construct a parametrized state $\rho_\lambda$ for $\lambda\in[0,1]$ in the form of a thermal state
\begin{equation}
    \begin{aligned}
	    &\rho_{\lambda} = \frac{1}{Z}  e^{- \beta (\lambda) H(\lambda)},
	    &Z = {\rm Tr}[e^{- \beta (\lambda) H(\lambda)}],
	\end{aligned}
\end{equation}
where $\beta(\lambda)$ is an effective inverse temperature satisfying conditions:
\begin{equation} \label{eq:beta_cond}
    \beta(0) \rightarrow \infty, \quad 
    \beta(1/2) = 1, \quad
    \beta(1) = 0,
\end{equation}
and $H(\lambda)$ is effective Hamiltonian, satisfying conditions:
\begin{equation}
    \quad H(1/2)=H_{\rm in}, \quad 
    {\rm g.s.}[H(0)]=\ket{\phi}\bra{\phi}.
\end{equation}
Here ${\rm g.s.}[H(0)]$ stands for the ground state of the Hamiltonian $H(1)$, and $\ket{\phi}$ is certain maximally entangled state.
The above construction ensures that
\begin{equation} \label{eq:conditions-on-rho}
    \rho_0=\ket{\phi}\bra{\phi},
    \quad 
    \rho_{1/2}=\rho_{\rm in}, 
    \quad
    \rho_{1}=d_A^{-1}d_B^{-1}{\bf 1}_A\otimes {\bf 1}_B.
\end{equation}
Next we use $\lambda$ as the confusion parameter.

To satisfy Eq.\eqref{eq:beta_cond}, we choose
\begin{equation}
    \beta(\lambda) = 1/\rm{tg} (\pi / (2 \lambda)).
\end{equation}
To construct $H(\lambda)$, we first represent $H_{\rm in}$ as a sum of two terms
\begin{equation}
    H_{\rm in} = H_{\rm in}^{\rm loc} + H_{\rm in}^{\text{non-loc}}
\end{equation}
corresponding to `local' and `non-local' parts given by
\begin{align}
       & H_{\rm in}^{\rm loc} = \sum_{i = 0}^{d^2_A-1} h_{i,0} \Lambda^{i}_A \otimes 
         \mathbf{1}_B + \sum_{i = 1}^{d^2_B-1} h_{0,j} \mathbf{1}_A \otimes \Lambda^{j}_B, \\
       & H_{\rm in}^{\text{non-loc}} = \sum_{i=1}^{d^2_A-1}\sum_{j = 1}^{d^2_B-1} h_{i, j} \Lambda^{i}_A 
         \otimes \Lambda^{j}_B,
\end{align}
where
\begin{equation}
    h_{i,j} = d_A^{-1}d_B^{-1}{\rm Tr}\left[H_{\rm in} \Lambda^i_A \otimes \Lambda^j_B\right],
\end{equation}
and $\{\Lambda^{i}_{A(B)}\}_{i=0}^{d^2_{A(B)}-1}$ are sets of Hermitian matrices satisfying conditions 
\begin{equation}
    {\rm Tr}[\Lambda_{A(B)}^i \Lambda_{A(B)}^j]=d_{A(B)}\delta_{i,j}, \quad
    \Lambda^0_{A(B)}={\bf 1}_{A(B)},
\end{equation}
i.e. generalized Gell-Mann matrices.
One can think about $H_{\rm in}^{\rm loc}$ as the term responsible for interacting with the local fields that cannot produce the entanglement. 
In contrast, the non-local part $H_{\rm in}^{\text{non-loc}}$ describes interaction between particles and thus can create an entanglement. 

Then we take $H(\lambda)$ in the form
\begin{equation}\label{eq:HamTransform}
       H (\lambda) = a(\lambda)H_{\rm in}^{\rm loc} + b(\lambda)
       \widetilde{U}(\lambda)H_{\rm in}^{\text{non-loc}} \widetilde{U}(
       \lambda)^{\dagger},
\end{equation}
where the smooth monotonic functions $a(\lambda)$ and $b(\lambda)$ serve as switching between local and non-local parts of Hamiltonian, 
and $\widetilde{U}(\lambda)$ is a specially-designed unitary operator whose aim to convert the ground state of $H_{\rm in}^{\text{non-loc}}$ into a maximally entangled state $\ket{\phi}$. 
The functions $a(\lambda)$ and $b(\lambda)$ have to satisfy the following conditions:
\begin{equation}
    \begin{aligned}
        &a(0) = b(1) = 0, \\
        &a(1)= b(0) \approx a(1/2) =b(1/2) \approx 1.
    \end{aligned}
\end{equation}
In our implementation, we consider these functions of the following form:
\begin{equation}
\begin{aligned}
    & a(\lambda) = 1 - \frac{2}{1 + \exp (\lambda / \delta)}, \\
    & b(\lambda) = \frac{2}{1 + \exp ((\lambda - 1) / \delta)} - 1,
\end{aligned}     
\end{equation}
where we fix $\delta=0.01$.
As we can see for $\lambda = 1/2$ the values of both functions are approximately equal to $\delta $ that is enough for our purposes.


Finally, we construct $\widetilde{U}(\lambda)$ in such a way that
\begin{equation}
    \widetilde{U}(1/2) = {\bf 1}_A\otimes {\bf 1}_B, \quad \widetilde{U}(1) \ket{g} = \ket{\phi},
\end{equation}
where $\ket{g}$ is a ground state of $H_{\rm in}^{\text{non-loc}}$, and, remember, $\ket{\phi}$ is some maximally entangled state.
Let 
\begin{equation}
       \ket{g} = \sum_{i} h_{i} \ket{\xi^A_{i}} \otimes \ket{\xi^B_{i}},
\end{equation}
for $h_i\geq 0$ be a Schmidt decomposition of $\ket{g}$ (i.e. $\{\ket{xi^A_{i}}\}_i$ and $\{\ket{xi^B_{i}}\}_i$ are taken from some orthonormal basis of $A$ and $B$ respectively).
We choose $\ket{\phi}$ in the form
\begin{equation}
    \ket{\phi} = \frac{1}{\sqrt{n}}  \sum_{i} 
         \ket{\xi^A_{i}} \otimes \ket{\xi^B_{i}},
\end{equation}
where $n=\min{(d_A,d_B)}$, providing minimal Hilbert-Schmidt distance between $\ket{g}$ and $\ket{\phi}$.

Then we construct $\widetilde{U}(\lambda)$ using its own effective Hamiltonian $\widetilde{H}$ and effective time $\Theta(\lambda)$:
\begin{equation}
    \widetilde{U}(\lambda) = \exp(-\imath \widetilde{H} \Theta(\lambda)).
\end{equation}
The effective time function $\Theta(\lambda)$ can be taken in the form
\begin{equation}
	\Theta(\lambda) = \frac{2}{1 + \exp(\lambda / \delta)},
\end{equation}
while the effective Hamiltonian $\widetilde{H}$ can be obtained as
\begin{equation}
    \widetilde{H} = \imath \ln [ \ket{\phi}\bra{g} + (\ldots) ],
\end{equation}
where $(\ldots)$ is arbitrary operator providing $\ket{\phi}\bra{g} + (\ldots)$ to be a unitary operator.

Taking everything together, we obtain state $\rho_\lambda$ satisfying conditions Eq.\eqref{eq:conditions-on-rho}.

\end{document}